\documentclass{article}
\usepackage{spconf,amsmath,graphicx}
\usepackage{subcaption}
\usepackage{comment}
\usepackage[unicode=true,colorlinks=true,linkcolor=magenta, urlcolor=blue, citecolor = blue,breaklinks]{hyperref}
\usepackage{enumitem}
\setlist{nosep, leftmargin=14pt}

\usepackage{mwe} 

\usepackage[dvipsnames]{xcolor}

\title{Hierarchical loss and geometric mask refinement for multilabel ribs segmentation}
\name{
\begin{tabular}{@{}c@{}}
Aleksei Leonov\sthanks{equal contribution} 
      \qquad Aleksei Zakharov$^{\ast}$ 
      \qquad Sergey Koshelev $^{\ast}$
      \\
    \textit{Maxim Pisov} 
    \qquad \textit{Anvar Kurmukov}
    \qquad \textit{Mikhail Belyaev} 
\end{tabular}
}

\address{IRA Labs Ltd} 

\begin{document}
%
\maketitle
\begin{abstract}
Automatic ribs segmentation and numeration can increase computed tomography assessment speed and reduce radiologists mistakes.
We introduce a model for multilabel ribs segmentation with hierarchical loss function, which enable to improve multilabel segmentation quality. 
Also we propose postprocessing technique to further increase labeling quality.
Our model achieved new state-of-the-art $98.2\%$ label accuracy on public RibSeg v2 dataset, surpassing previous result by $6.7\%$.
\end{abstract}

\begin{keywords}
rib segmentation, rib labeling, computed tomography
\end{keywords}
\section{Introduction}
Rib fractures represent the most frequently encountered pathology within thoracic traumas, and the ensuing complications can inflict significant anguish upon patients    \cite{fracs_info_new}.
For each fracture, the radiologist have to determine the rib number and location, which is a time-consuming procedure because the ribs are elongated and their type cannot be determined on a single axial slice (see Fig.\ref{fig:back_mask}, the first column).
Automatic segmentation and numeration of ribs can significantly simplify and speed up the analysis of rib pathologies, and also reduce radiologists' mistakes.

Several deep learning algorithms for ribs segmentation have appeared in recent years (see Section \ref{section:related}). 
Most of them use internal datasets not available for public use or review. 
To address these challenges, the public dataset RibSeg v2 \cite{RibSeg_v2} was released. 
The authors compared nnU-Net \cite{nnUnet} with several point cloud models and showed that point cloud model DGCNN \cite{DGCNN} achieved better segmentation and labeling results, although having problems with detection of first ribs and processing challenging cases (see Tabs. \ref{'tab:RS AFIT metrics'}, \ref{tab:challenging}).
Using this open source benchmark for training and evaluation, we introduce a new model that achieved state-of-the-art results on the RibSeg v2 dataset, received $98.2\%$ label accuracy ($+6.7\%$ to \cite{RibSeg_v2}). 
Also, proposed model is much more robust and reached $97.4\%$ label accuracy ($+17.8\%$ to \cite{RibSeg_v2}) for challenging cases.

\noindent\textbf{Our contribution} is two-fold:
\begin{itemize}
    \item 
    First, we propose hierarchical loss function for ribs simultaneous segmentation and labeling.
    Our model is one-stage and trained end-to-end, which increase training and inference speed and accuracy. 
    \item 
    Second, we suggest simple postprocessing technique called geometric mask refinement, which can be used to improve ribs labeling accuracy.

\end{itemize}
\noindent
Futhermore, we conduct an audit of RibSeg v2 dataset and highlight problems in annotations which are necessary to rectify for fair models' quality assessment and comparison. Moreover, rectification of these inaccuracies can be usefull for training better models.

\section{Related Work}
\label{section:related}

Prior to the widespread adoption of deep learning in medical imaging, several studies explored rib segmentation and labeling with a variety of methods such as  using image primitives \cite{old_image_primitives} and ray search \cite{old_ray_search}. 
Nevertheless, these methods were highly susceptible to local ambiguities and seed points.

All latest methods are deep learning-based, with U-Net \cite{Unet} being the standard architecture.
nnU-Net framework \cite{nnUnet} for 3D medical images segmentation is widely used due to its ability to adapt to specific task data.

Multilabel ribs segmentation is usually decomposed into two steps: binary ribs segmentation and consecutive connected components processing.
In various studies rib labeling is realized as a counting like growing regions process \cite{region_growing} or connected domain algorithm \cite{connected_domain_algorithm}.
Additionally, the study in \cite{backbone_labeling} employs a method of labeling ribs based on the nearest thoracic vertebra number.
Work \cite{Fisrt_Inter_Twelfth} labels ribs by using the classification of the 1st, 12th and intermediate ribs. 
All these methods may show good results when ribs are easily separable and each  rib consists of one component, which may not coincide with reality when rib is broken or when several ribs are connected together. Besides these algorithms fail if ribs binary segmentation is imperfect.

Direct segmentation of 24 ribs is implemented in TotalSegmentator \cite{TS} and RibSeg v2 \cite{RibSeg_v2} pipelines. Authors of TotalSegmentator use nnUnet framework to segment 104 anatomical structures including ribs.
However, a significant portion of the images in TotalSegmentator is from the abdominal area. Of the 65 total test images, only 26 images are of the thorax. Also ribs detection metrics are not provided.

Therefore, for training and validation of our algorithm we use the RibSeg v2 dataset (see Section \ref{section:data}), which is created especially for ribs segmentation and labeling.
The authors of RibSeg v2 use a two-step pipeline: first network classifies  voxels to foreground (ribs) and background, while second network classifies foreground into 24 classes.
They employ point-based network architectures, selecting DGCNN \cite{DGCNN} model as the best solution.
This model leverages local geometric structures by constructing a neighborhood graph and applying convolution-like operations on the graph edges.




\section{Method}

\subsection{Architecture}
We utilize a standard U-Net architecture  \cite{Unet} with 5 levels as a backbone.
The first level contains 16 channels, with the number of channels doubling at each subsequent level. 
We use ResBlocks everywhere except first convolution and Dilated Fusion block \cite{DilFusion} on the last level.

There are two heads on the output - binary and classification.
The binary head has one channel and is used for ribs' binary segmentation, 
while the classification head has 12 channels, one for each rib type (1-12).
This division into two heads is crucial as it enables the training of the classification head specifically in the area where ribs are located.

\subsection{Hierarchical loss function}
Our loss function consists of binary and multi-label parts.
Binary loss is responsible for ribs binary segmentation to separate ribs from background, while ribs classification is trained by multi-label loss. 
Such separation allows us to make loss hierarchical and penalize classification head just in voxels with ribs.
This technique helps to converge faster and enables to use common cross-entropy loss function for classification, because ribs classes became balanced. 
Also due to such technique we can train our network end-to-end which improves train and inference speed and complexity.

\begin{equation}
\label{eq:loss}
\begin{split}
Loss = & (Dice + Focal + BCE)[t^{bin}, p^{bin}] + \\
       & \alpha \cdot \sum_{j | t^{bin}_j = 1} (CE + SAM) [t^{cls}_j, p^{cls}_j]
\end{split}
\end{equation}

Binary loss itself consists of three components: $Dice$ , $Focal$, and $BCE$ (Binary Cross Entropy) losses.
This combination is important because even for binary ribs segmentation foreground and background classes are imbalanced (ribs usually occupy less than $0.5\%$ of an image).

Multi-label loss in turn consists of two components: $CE$ and $SoftArgMax$ loss \cite{SAM}.
$SoftArgMax$ loss is a differentiable relaxation of $argmax$ function.
This loss can be used for this task because ribs classes have a strict order.
While cross entropy penalize every wrong class with the same magnitude, $SoftArgMax$ loss increases when predicted class is more distant from ground truth class.

In equation \ref{eq:loss} ribs binary segmentation target denoted as $t^{bin}$, ribs classes target as $t^{cls}$,
network binary and classification heads outputs as $p^{bin}$ and $p^{cls}$.
We used $\alpha = 0.05$ in final experiments.

\subsection{Geometric mask refinement}
\label{section:postprocessing}
As we show in Section \ref{section:results}, lower ribs are harder to classify than upper ones.
Therefore, after primary filtration from small components,
we add geometric mask refinement process to our network to improve lower ribs classification. The algorithm is illustrated in Fig.\ref{fig:process}:

\begin{figure}[t] 
    \centering
        \includegraphics[width=1.\linewidth]
        {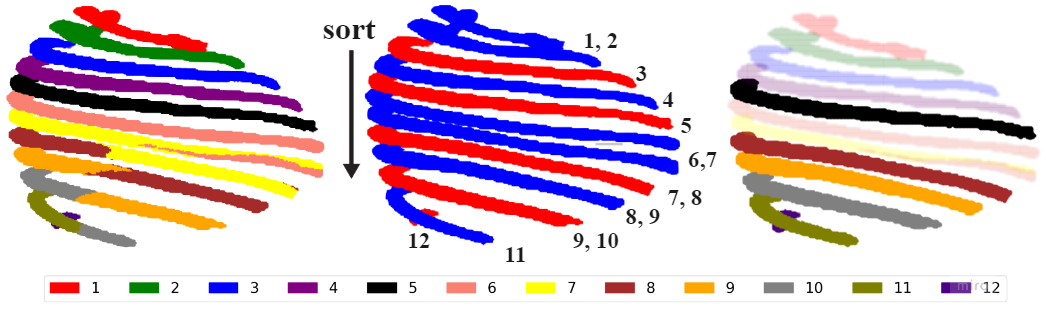}
        \caption{\textbf{Postprocessing algorithm.} For simplicity, only right ribs are shown. Left: initial prediction. Middle: connected components are shown with alternating red and blue. Numbers indicate "probable" ribs types. Right: final result. Untouched components are transparent.}
        \label{fig:process}    
\end{figure}

\begin{enumerate}
    \item Separate right and left ribs and split ribs into components by predicted ribs binary mask.
    \item In each group sort the ribs components according to their height.
    \item For each component calculate how many ribs could be in this component. 
    Each component may include the amount of ribs equal to the number of times it is larger than the median volume of all components.
    This step is important because some ribs may be connected.
    \item Calculate "probable" rib types for each component.
    Rib type considered "probable" if voxels of this type occupy more than  a third of component volume. (Fig.\ref{fig:process}, middle)
    
    \item Choose sequence of components types that include largest amount of "probable" types among all possible consecutive sequences.
    If top visible ribs on the image are the first ones (ribs 1-4) they stay untouched and sequence is chosen just for the next components. (Fig.\ref{fig:process}, right)
\end{enumerate}
This postprocessing technique is highly effective and robust, 
enhancing or maintaining the quality of nearly every prediction, as detailed in Tabs. \ref{'tab:RS AFIT metrics'} and \ref{table:comparison}.

\section{Data and Metrics}
\label{section:data}

\subsection{RibSeg v2 Dataset}
\label{section:ribseg}
The RibSeg v2 dataset \cite{RibSeg_v2} is specifically created for ribs segmentation, labeling, and center-line extraction. 
It comprises a total of 660 CT scans sourced from the RibFrac dataset \cite{ribfrac2020} designed for rib fractures segmentation.
We used the same data split as the original study: 420 CT scans for train, 80 scans for evaluation (6 unqualified due to missing or flawed annotations), and 160 scans for test.

For making ground truth annotation, the authors used two approaches: morphological-based and centerline-based.
The first faster approach was used in most cases, while the second was applied in cases where first method failed. 
Both approaches are semi-automatic, which as we have found led to poorer annotation quality.
The most significant issues are observed in the regions near the vertebrae.
As seen on Fig. \ref{fig:back_mask}, some cases have extra ribs segmentation, while some others have significant cutting in segmentation near to the spine.
Such annotation lead to worse model quality and also could significantly decrease test metrics.

\subsection{Metrics}
\label{sec:metrics}
For fair comparison, we use metrics which are provided by the authors of RibSeg v2.
Ribs' detection quality for clinical applicability is assessed by the Label-Accuracy of individual ribs. 
Specifically, an individual rib $i$ is counted as correctly labeled if $Recall_{i} > 0.7$. 
Taking into account the characteristic difference between the first and twelfth ribs in shape and elongation, the authors separately consider the Label-Accuracy value for all (A) / first (F) / intermediate (I) / twelfth (T) rib pairs respectively.

For evaluation of segmentation quality, the authors of the RibSeg v2 used Label-Dice of the 24 ribs, 
denoted as $Dice^{(L)}_i = \frac{2 \cdot |y_i \cap \hat{y}_i|}{|y_i| + |\hat{y}_i|}$,
where $y$ and $\hat{y}$ indicate the label prediction and ground truth respectively.
If the ground truth mask is empty for a rib, then the Dice Score is not used for scoring.
The authors evaluate segmentation performance with average Dice Score $Dice^{L}_{\text{avg}}$ and report performance degradation with minimal Label-Dice among all ribs, denoted as $Dice^{L}_{\text{min}}$.


\section{Experimental setup}

\begin{figure}[!t]
    \centering

    \begin{subfigure}{0.32\columnwidth}
        \includegraphics[height=2.3cm,width=\linewidth]{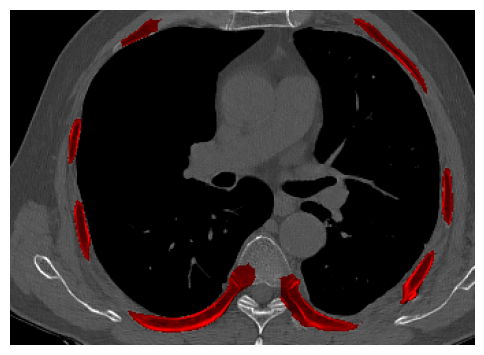}
    \end{subfigure}
    \begin{subfigure}{0.32\columnwidth}
        \includegraphics[height=2.3cm,width=\linewidth]{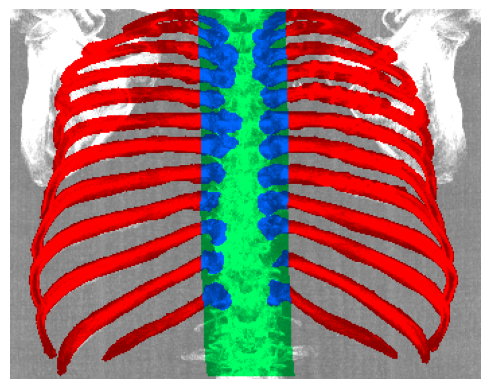}
    \end{subfigure}
    \begin{subfigure}{0.32\columnwidth}
        \includegraphics[height=2.3cm,width=\linewidth]{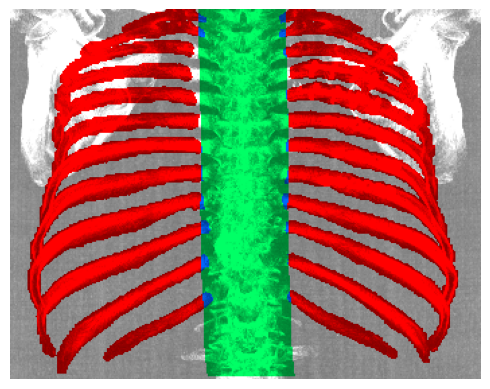}
    \end{subfigure}
    \\ 
    \begin{subfigure}{0.32\columnwidth}
        \includegraphics[height=2.3cm, width=\linewidth]{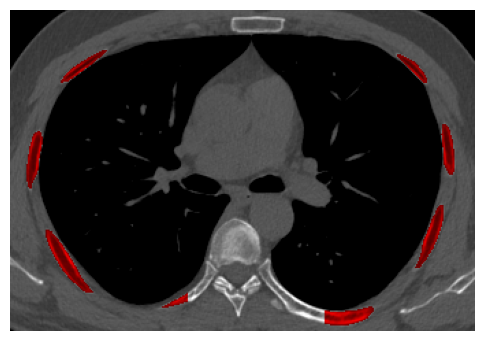}
        \caption*{annotation on axial}
    \end{subfigure}
    \begin{subfigure}{0.32\columnwidth}
        \includegraphics[height=2.3cm, width=\linewidth]{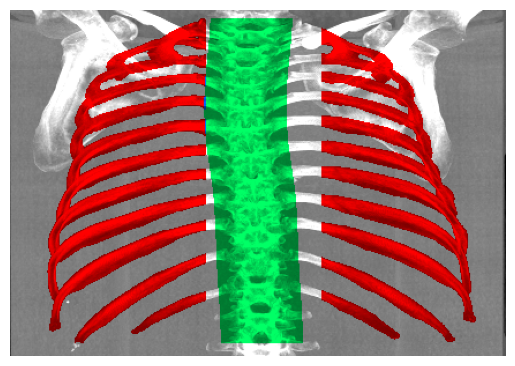}
        \caption*{annotation}
    \end{subfigure}
    \begin{subfigure}{0.32\columnwidth}
        \includegraphics[height=2.3cm, width=\linewidth]{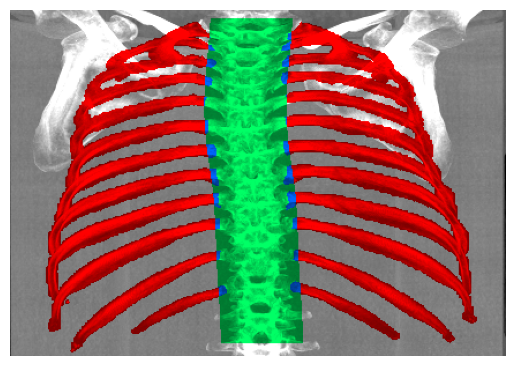}
        \caption*{prediction}
    \end{subfigure}
    
    \caption{\textbf{Filtration of the ribs near the spine}. Center and right columns: coronal plain - annotation and prediction masks with filtering respectively. Red color indicates ribs mask, green color shows area where we filter masks to mitigate the impact of inaccurate annotation, blue color denotes filtered part of ribs.}
    \label{fig:back_mask}
\end{figure}

\subsection{Pipeline}\label{sec:pipeline}
At the preprocessing stage, each image is linearly zoomed to achieve a uniform resolution of 2 mm along each axis. 
Then we do min-max normalization using bone window -- from $-450$ to $1050$ Hounsfield Units (HU) values. 
Pixel values below -450 are set to 0, and those exceeding 1050 are set to 1.
Training and inference were conducted  on a machine equipped with a single NVIDIA A100 Tensor Core GPU with 40G of memory.

\subsection{Training}
The model was trained for 4800 iterations with batch size of 8, using ADAM optimizer with automatic mixed precision mode.
%
A cosine scheduler with an initial linear warmup phase for the first 640 iterations was employed. 
Various data augmentations were applied, including flips, rotations on all axes, padding and cropping.
Finally, random patches of uniform size 320 mm along the transverse (vertical) axis were generated.
This ensures the model is trained on varied portions of the CT scan, further enhancing its ability to generalize across different scans and scenarios.

\subsection{Inference}
During the inference phase, preprocessed image (see Section \ref{sec:pipeline}) is divided into patches of uniform size along the vertical axis, as in the training phase. 
These patches are chosen such that adjacent ones overlap by half the patch size. 
For each patch, the output from the first head is transformed using a sigmoid function to produce a prediction of binary probabilities.
To generate a prediction of classification probabilities, the output from the second head undergoes a softmax transformation.
Within overlapping regions the predictions are averaged.
Ribs binary segmentation mask emerges by thresholding the binary prediction at $0.25$ (based on binary segmentation metrics on validation dataset).
To classify voxels in binary mask, the peak value across the 12 channels is chosen from classification head prediction.

\section{Results}
\label{section:results}

We compare our results with the DGCNN \cite{DGCNN}, chosen by authors of RibSeg v2 as the best point-based model, and nnU-Net \cite{nnUnet}, which was presented as a SOTA voxel-based model.





\begin{table}[!t]
\caption{\textbf{Ribs segmentation metrics on RibSeg v2 Test Set}.
\textit{Our} - initial model without post-processing; \\
Manual nnU-Net - manually trained  nnU-Net with 8G (as in RibSeg v2) and 24G GPU memory at training. \\
+Proc - with post-processing; * - with cut near the spine. \\
A - all, F - first, I - intermediate, T - twelfth rib pairs.
}

\centering
\begin{tabular}{lccccc}
\hline
& \multicolumn{4}{c}{Label-Accuracy} & $Dice^{(L)}_{\text{avg}}$ \\
 & A & F & I & T &  \\

\hline
nnU-Net & 87.5 & 92.9 & 87.7 & 78.5  & 83.6 \\
DGCNN  & 91.5 & 60.7 & 96.3 & 89.4 & 90.5 \\

\textit{Our} & 92.6 & 92.3 & 93.5 & 83.0  & 78.8 \\
\textit{Our+Proc}  & \textbf{94.1} & 92.3 & 95.2 & 82.3  & 79.4\\
\textit{Our*}  & 96.6 & 99.0 & 96.7 & 92.1 & 82.5 \\
\textit{Our+Proc*}  & \textbf{98.2} & 99.0 & 98.7 & 91.7 & 83.7 \\

\hline
\multicolumn{6}{l}{Manual nnU-Net} \\

$\text{8G}$ & 79.8 & 80.6 & 80.1 & 75.7  & 83.0 \\
$\text{24G}$ &  86.2 &  83.9 &  87.4 &  74.5 & 87.7 \\
$\text{24G}\text{+Proc}$ &  88.6 &  83.9 &  90.3 &  74.1 &  89.3\\
$\text{24G*}$  &  89.3 &  98.1 &  89.6 &  76.4 & 88.7 \\
$\text{24G}\text{+Proc*}$  &  92.0 &  98.1 &  92.7 &  76.0  & 90.3 \\
\hline
\end{tabular}
\label{'tab:RS AFIT metrics'}
\end{table}

Main results are represented in Tab. \ref{'tab:RS AFIT metrics'}. 
Initial model without postprocessing called \textit{Our}  achieved 92.6\% label-accuracy, outperforming DGCNN model by 1\% and significantly outperforming voxel-based nnU-Net (by 5\%) but has worse dice score. 
This may indicate that model is better in ribs labeling, while binary segmentation is more coarse.

We implemented postprocessing (see Section \ref{section:postprocessing}) after identifying challenges in the network's classification of lower ribs (ribs 7-11), as evidenced by Table \ref{table:comparison}.
It could be due to suboptimal architecture or training process.
Perfect models need to capture the full image context, as the structure of ribs varies between individuals, from 11 to even 13.
Table \ref{table:comparison} shows postprocessing improvement for each rib type.
Lower ribs numeration degradation can be seen along with effect of postprocessing algorithm up to 8\% and 7\% gain in label-accuracy for 10th and 11th ribs respectively.
This improvement allows us to reach 94.1\% average label-accuracy for all ribs.

One more difficulty that we faced was inaccurate annotation mentioned in Section \ref{section:ribseg}.
Due to target errors, some good predictions were counted as wrong ones. 
As can be seen on the Fig. \ref{fig:back_mask}, inaccuracies in annotations can seriously affect segmentation metrics even for visually perfect predictions. 
Therefore, for a fairer assessment of quality, we prioritize detection metrics. Moreover, we also calculated metrics with the target and prediction cut off at a distance of 30 mm from the spine center line \cite{backline} (\textit{Our*} and \textit{Our+Proc*} in Tab. \ref{'tab:RS AFIT metrics'}). 
This transformation only affects the area in the immediate vicinity of the spine, where target is incorrect for many cases.
Using this validation procedure, our model achieves 98.2\% average label-accuracy for all ribs.
There are just a couple of 160 test cases where model makes serious mistakes.

In order to study the robustness of the geometric mask refinement, we trained nnU-Net with a configuration similar to that proposed by the authors of RibSeg v2 (8G in Tab.\ref{'tab:RS AFIT metrics'}). The average dice was comparable to published metrics, but the detection metrics were significantly lower due to unknown factors. To ensure a fair comparison between the proposed model and nnU-Net, we also trained nnU-Net using 24G GPU memory. Nonetheless, the comparison of the metrics demonstrate the effectiveness of the proposed postprocessing, increasing label-accuracy by 2-3\%.

Additionally, the RibSeg v2 authors reported metrics for \textit{normal} and \textit{challenging} groups of studies. 
They visually selected 99 cases as \textit{challenging} (47/19/33 in training/evaluation/test sets). Such cases present segmentation and classification difficulties due to factors like connected adjacent bones, partial scans, pronounced bone lesions, scoliosis, fractures, indistinct or absent floating ribs, and presence of a third short 13th floating rib.

\setlength{\tabcolsep}{3pt} 
\begin{table}[!t]
\caption{\textbf{Label-Accuracy comparison for ribs types and postprocessing effect}. Ribs from 2 to 5 merged to one cell due to the proximity of values.}
\centering
\begin{tabular}{lccccccccc}
\hline
& \textbf{1} & \textbf{2-5} & \textbf{6} & \textbf{7} & \textbf{8} & \textbf{9} & \textbf{10} & \textbf{11} & \textbf{12} \\
\hline
\textit{Our*} & 99.0 & ~99.5 & 98.4 & 97.4 & 96.8 & 95.2 & 90.6 & 89.5 & 92.1 \\
\textit{+Proc} & 99.0 & ~99.5 & 99.0 & 98.7 & 98.7 & 98.1 & 97.7 & 96.1 & 91.7 \\
$\Delta$ & 0.0 & 0.0 & \textcolor{ForestGreen}{+0.6} & \textcolor{ForestGreen}{+1.3} & \textcolor{ForestGreen}{+1.9} & \textcolor{ForestGreen}{+2.9} & \textcolor{ForestGreen}{+7.1} & \textcolor{ForestGreen}{+6.6} & \textcolor{red}{-0.4} \\
\hline
\end{tabular}
\label{table:comparison}
\end{table}

\setlength{\tabcolsep}{2pt} 
\begin{table}[!t]
    \centering
    \caption{\textbf{Models comparison on  normal/challenging cases}. Cases taken from RibSeg v2 test set. Label-Accuracy  (Acc) counted over all pairs of ribs.}
    \begin{tabular}{lcccccc}
        \hline
        & \multicolumn{3}{c}{Normal} & \multicolumn{3}{c}{Challenging} \\
          & Acc & $Dice^{L}_{\text{avg}}$ & $Dice^{L}_{\text{min}}$  & Acc & $Dice^{L}_{\text{avg}}$ & $Dice^{L}_{\text{min}}$ \\
        \hline
        DGCNN & 94.0 & 96.2 & 73.4  & 79.6 & 72.3 & 62.4  \\
        \textit{Our+Proc} & \textbf{95.2} & 80.1 &  63.7 & \textbf{89.2} & 76.6 &  44.6   \\
        \textit{Our+Proc*} & \textbf{98.4} & 83.8 &  72.2  & \textbf{97.4} & 83.5 &  58.4  \\
        \hline
    \end{tabular}
    \label{tab:challenging}
\end{table}

The results, as demonstrated in Tab. \ref{tab:challenging}, indicate a significantly enhanced robustness of our model.
There is almost no drop in average label-accuracy. 
We get 89\% (97\% for cut targets) label-accuracy which is 10\% (20\%) higher than RibSeg v2 authors results on \textit{challenging} cases.

\section{Conclusion}

We presented new deep learning algorithm for ribs segmentation and labeling. 
Our algorithm achieved $98.2\%$ label-accuracy on the public RibSeg v2 dataset.
Such results make it possible to use this algorithm in radiologists routine work.
Some mistakes still occur, but they can be mitigated with more diverse training data and more accurate annotations.



\section{Compliance with Ethical Standards}
This research study was conducted retrospectively using  public datasets. Ethical approval was not required as confirmed by the license attached with the open access data.

\section{Conflicts of Interests}    
No funding was received for conducting this study. The authors have no relevant financial or non-financial interests to disclose.


\bibliographystyle{IEEEbib}
\bibliography{strings,refs}

\begin{thebibliography}{10}

\bibitem{fracs_info_new}
Brett~S. Talbot, Christopher~P. Gange, Apeksha Chaturvedi, Nina Klionsky, Susan~K. Hobbs, and Abhishek Chaturvedi,
\newblock ``Traumatic rib injury: Patterns, imaging pitfalls, complications, and treatment,''
\newblock {\em RadioGraphics}, vol. 37, no. 2, pp. 628--651, 2017,
\newblock PMID: 28186860.

\bibitem{RibSeg_v2}
Liang Jin, Shixuan Gu, Donglai Wei, Jason~Ken Adhinarta, Kaiming Kuang, Yongjie~Jessica Zhang, Hanspeter Pfister, Bingbing Ni, Jiancheng Yang, and Ming Li,
\newblock ``Ribseg v2: A large-scale benchmark for rib labeling and anatomical centerline extraction,'' 2023.

\bibitem{nnUnet}
Fabian Isensee, Jens Petersen, Andre Klein, David Zimmerer, Paul~F. Jaeger, Simon Kohl, Jakob Wasserthal, Gregor Koehler, Tobias Norajitra, Sebastian Wirkert, and Klaus~H. Maier-Hein,
\newblock ``nnu-net: Self-adapting framework for u-net-based medical image segmentation,'' 2018.

\bibitem{DGCNN}
Yue Wang, Yongbin Sun, Ziwei Liu, Sanjay~E. Sarma, Michael~M. Bronstein, and Justin~M. Solomon,
\newblock ``Dynamic graph cnn for learning on point clouds,'' 2019.

\bibitem{old_image_primitives}
Joes Staal, Bram {van Ginneken}, and Max~A. Viergever,
\newblock ``Automatic rib segmentation and labeling in computed tomography scans using a general framework for detection, recognition and segmentation of objects in volumetric data,''
\newblock {\em Medical Image Analysis}, vol. 11, no. 1, pp. 35--46, 2007.

\bibitem{old_ray_search}
Tobias Klinder, Cristian Lorenz, Jens von Berg, Sebastian P.~M. Dries, Thomas B{\"u}low, and J{\"o}rn Ostermann,
\newblock ``Automated model-based rib cage segmentation and labeling in ct images,''
\newblock in {\em Medical Image Computing and Computer-Assisted Intervention -- MICCAI 2007}, Nicholas Ayache, S{\'e}bastien Ourselin, and Anthony Maeder, Eds., Berlin, Heidelberg, 2007, pp. 195--202, Springer Berlin Heidelberg.

\bibitem{Unet}
Olaf Ronneberger, Philipp Fischer, and Thomas Brox,
\newblock ``U-net: Convolutional networks for biomedical image segmentation,'' 2015.

\bibitem{region_growing}
Yu~Jin Seol, So~Hyun Park, Young~Jae Kim, Young-Taek Park, Hee~Young Lee, and Kwang~Gi Kim,
\newblock ``The development of an automatic rib sequence labeling system on axial computed tomography images with 3-dimensional region growing,''
\newblock {\em Sensors}, vol. 22, no. 12, 2022.

\bibitem{connected_domain_algorithm}
Mingxiang Wu, Zhizhong Chai, Guangwu Qian, Huangjing Lin, Qiong Wang, Liansheng Wang, and Hao Chen,
\newblock ``Development and evaluation of a deep learning algorithm for rib segmentation and fracture detection from multicenter chest ct images,''
\newblock {\em Radiology: Artificial Intelligence}, vol. 3, no. 5, pp. e200248, 2021.

\bibitem{backbone_labeling}
Zhen wei Lin, Wei li~Dai, Qing-Quan Lai, and Hong Wu,
\newblock ``Deep learning-based computed tomography applied to the diagnosis of rib fractures,''
\newblock {\em Journal of Radiation Research and Applied Sciences}, vol. 16, no. 2, pp. 100558, 2023.

\bibitem{Fisrt_Inter_Twelfth}
Matthias Lenga, Tobias Klinder, Christian Bürger, Jens von Berg, Astrid Franz, and Cristian Lorenz,
\newblock ``Deep learning based rib centerline extraction and labeling,''
\newblock in {\em Computational Methods and Clinical Applications in Musculoskeletal Imaging}, pp. 99--113. Springer International Publishing, 2019.

\bibitem{TS}
Jakob Wasserthal, Hanns-Christian Breit, Manfred~T. Meyer, Maurice Pradella, Daniel Hinck, Alexander~W. Sauter, Tobias Heye, Daniel~T. Boll, Joshy Cyriac, Shan Yang, Michael Bach, and Martin Segeroth,
\newblock ``{TotalSegmentator}: Robust segmentation of 104 anatomic structures in {CT} images,''
\newblock {\em Radiology: Artificial Intelligence}, vol. 5, no. 5, sep 2023.

\bibitem{DilFusion}
Jiahang Wang, Wei Zhang, Weizhong Liu, and Tao Mei,
\newblock ``Down to the last detail: Virtual try-on with detail carving,'' 2020.

\bibitem{SAM}
Diogo~C. Luvizon, Hedi Tabia, and David Picard,
\newblock ``Human pose regression by combining indirect part detection and contextual information,'' 2017.

\bibitem{ribfrac2020}
Liang Jin, Jiancheng Yang, Kaiming Kuang, Bingbing Ni, Yiyi Gao, Yingli Sun, Pan Gao, Weiling Ma, Mingyu Tan, Hui Kang, Jiajun Chen, and Ming Li,
\newblock ``Deep-learning-assisted detection and segmentation of rib fractures from ct scans: Development and validation of fracnet,''
\newblock {\em EBioMedicine}, vol. 62, pp. 103106, 12 2020.

\bibitem{backline}
Alexey Zakharov, Maxim Pisov, Alim Bukharaev, Alexey Petraikin, Sergey Morozov, Victor Gombolevskiy, and Mikhail Belyaev,
\newblock ``Interpretable vertebral fracture quantification via anchor-free landmarks localization,'' 2022.

\end{thebibliography}
\end{document}